\documentclass[footinbib,a4paper,aps,superscriptaddress,reprint,twocolumn,preprintnumbers,amsmath,amssymb,nobalancelastpage,10pt]{revtex4-1}
\usepackage[english]{babel}
\usepackage{amsmath}
\usepackage{verbatim}
\usepackage{graphicx}
\usepackage{color}
\usepackage{tikz}
\usepackage{siunitx}
\usepackage{physics}
\usepackage{subfigure}
\usepackage{float}

\usepackage{mathptmx}
\usepackage{amssymb}
\usepackage{amsmath}
\usepackage{amsfonts}
\usepackage{array}

\usepackage{bm}
\usepackage{xcolor}
\usepackage{braket}

\definecolor{mygrey}{gray}{0.35}
\definecolor{myblue}{rgb}{0.2,0.2,0.8}
\definecolor{myzard}{cmyk}{0,0,0.05,0}
\definecolor{mywhite}{rgb}{1,1,1}
\definecolor{myred}{rgb}{1,0.,0.3}

\usepackage[colorlinks=true,citecolor=myblue,linkcolor=myred]{hyperref}

\def\beq{\begin{equation}}
\def\eeq{\end{equation}}

\def\barray{\begin{eqnarray}}
\def\earray{\end{eqnarray}}

\begin{document}


\title{Higher-order topological quantum paramagnets}

\author{Daniel Gonz\'{a}lez-Cuadra}\email{daniel.gonzalez@icfo.eu}
\affiliation{ICFO - Institut de Ci\`encies Fot\`oniques, The Barcelona Institute of Science and Technology, Av. Carl Friedrich Gauss 3, 08860 Castelldefels (Barcelona), Spain}

\begin{abstract}
Quantum paramagnets are strongly-correlated phases of matter where competing interactions frustrate magnetic order down to zero temperature. In certain cases, quantum fluctuations induce instead topological order, supporting, in particular, fractionalized quasi-particle excitations. In this work, we investigate paradigmatic spin models and show how magnetic frustration can also give rise to higher-order topological properties. We first study the frustrated Heisenberg model in a square lattice, where a plaquette valence bond solid appears through the spontaneous breaking of translational invariance. Despite the amount of effort that has been devoted to study this phase, its topological nature has so far been overlooked. By means of tensor network simulations, we establish how such state belongs to a higher-order symmetry-protected topological phase, where long-range plaquette order and non-trivial topology coexist. This interplay allows the system to support excitations that would be absent otherwise, such as corner-like states in the bulk attached to dynamical topological defects. Finally, we demonstrate how this higher-order topological quantum paramagnet can also be induced by dipolar interactions, indicating the possibility to directly  observe this phase using atomic quantum simulators.
\end{abstract}

\maketitle

\paragraph*{Introduction.--} Topology has emerged in the last decades as a central concept in theoretical physics, acting as a driving force in the search of novel phases of matter~\cite{Moore_2010}. The growing interest in topological quantum states experienced in the last years has been fueled, in particular, by the outstanding experimental progress in ultracold atomic experiments~\cite{Jaksch_2005, Lewenstein_2007, Gross_2017, Schafer_2020}, where such states can be prepared and investigated with an unprecedented degree of control~\cite{Atala_2013, Hirokazu_2013, Aidelsburger_2013, Jotzu_2014, Aidelsburger_2015, Goldman_2016}, as well as by their potential applications in fault-tolerant quantum computation~\cite{Kitaev_2003, Dai_2017, Semeghini_2021}.

Although non-trivial topology can exist in the absence of any symmetry~\cite{Wen_2017}, both notions are usually intertwined. This is the case for symmetry-protected topological (SPT) phases, recently realized in atomic platforms~\cite{deLeseleuc_2019, Sompet_2021}, which are only robust to perturbations that preserve certain symmetries. In the last years, there has been a growing interest in the study of topological crystalline insulators (TCI)~\cite{Fu_2011}, which are not protected by global~\cite{Chiu_2016} but rather point-like symmetries, such as inversion or rotational invariance~\cite{Hughes_2011}. Within this class, the recently discovered higher-order topological insulators (HOTIs)~\cite{Benalcazar_2017, Benalcaza_2017_2} present unusual properties that generalize the standard bulk-boundary correspondence~\cite{Ryu_2002}: they support gapless states in boundaries of co-dimension larger that one, such as corner or hinge states~\cite{Schindler_2018, Khalaf_2018}. Although non-interacting HOTIs have been extensively investigated, and in several cases realized experimentally~\cite{Imhof_2018, Peterson_2018, Serra-Garcia_2018, Ni_2019, Xue_2019, Xue_2019_2, Kempkes_2019, Weiner_2020}, interacting higher-order symmetry-protected topological (HOSPT) phases are only starting to be explored~\cite{You_2018, Dubinkin_2019, Kudo_2019, Sil_2020, Rasmussen_2020, Bibo_2020, Peng_2020, Hackenbroich_2021, Otsuka_2021}.

In this work, we present a novel interaction-induced HOSPT phase driven by magnetic frustration: a higher-order topological quantum paramagnet (HOTQP). Quantum paramagnets (QPs) arise in spin models where competing interactions enhance quantum fluctuations and prevent standard magnetic order~\cite{Lacroix_2011}. Of particular interest are spin liquids~\cite{Savary_2016, Zhou_2017}, which do not break any symmetry and can present topological order. The latter usually compete with other QPs such as valence-bond solids (VBS), that spontaneously break certain spatial symmetries. Here we show how the plaquette VBS (PVBS) that appears in frustrated Heisenberg models belongs to a HOSPT phase. To the best of our knowledge, such HOTQP represents the first example of a a quantum phase that presents both long-range order (LRO) due to the spontaneous symmetry breaking (SSB) and non-trivial higher-order topological properties. We show how the interplay between these two features gives rise to new strongly-correlated topological effects. In particular, topological defects in the plaquette order are spontaneously created when moving away from the zero magnetization sector. These defects separate regions with different bulk topology and, as a consequence, they host corner-like states in the bulk. Although topological defects have been investigated before in HOTIs, previous studies have focused on static defects imposed externally~\cite{Benalcazar_2019, Liu_2019, Li_2020}. Our results show, however, how dynamical defects bound to localized states emerge in a translational invariant model.

The paper is organized as follows. We first consider the Heisenberg model in a square lattice in the presence of interactions beyond nearest neighbors (NN) and study a phase transition between a magnetically-ordered and a PVBS phase. We then show how the latter belongs to a HOTQP phase by revealing its entanglement structure, as well as by computing a many-body topological invariant. These quantities are also used to characterize the topological phase transition between the trivial and the HOTQP phases, and to investigate the mechanism behind the emergence of topological defects. Finally, we study the Heisenberg model with dipolar interactions and show that its ground state is also a HOTQP. The dipolar Heisenberg model has gained attention in the last years~\cite{Zou_2017, Keles_2018a, Yao_2018, Keles_2018b}, since it naturally appears as an effective description of ultracold molecules trapped in optical lattices~\cite{Baranov_2012, Bohn_2017}, and can also be simulated using trapped ions~\cite{Porras_2011, Grass_2014, Bermudez_2017, Monroe_2021}. Ultracold molecules are particularly interesting for quantum simulation purposes due to their strong dipole interactions, and different schemes have been proposed to use them to simulate quantum magnetism~\cite{Micheli_2006, Barnett_2006, Gorshkov_2011a, Gorshkov_2011b} as well as topological phases~\cite{Yao_2013, Manmana_2013, Gorshkov_2013, Peter_2015}. Recent experimental progress have demonstrated how to cool, trap~\cite{Ni_2008, Deiglmayr_2008, Ni_2010, Park_2015, DeMarco_2019} and control interactions between molecules~\cite{Chotia_2012, Yan_2013, Hazzard_2014}, indicating how the observation of a HOTQP could be within reach using near-term devices.

\vspace{0.5ex}
\paragraph*{The frustrated Heisenberg model.--} We start by considering the antiferromagnetic Heisenberg model on a two-dimensional square lattice,
\begin{equation}
\label{eq:model}
\hat{H} = \sum_{i, j} J_{i,j} \, \hat{\mathbf{S}}_i \cdot \hat{\mathbf{S}}_j,
\end{equation}
with $J_{i,j}>0$, where $\hat{\mathbf{S}}_i = (S^x_i, S^y_i, S^z_i)$ and $S^\mu_i = \frac{1}{2} \sigma^\mu_i$ are the usual spin-$1/2$ operators and $\sigma^\mu_i$ are Pauli matrices. If only NN interactions are present, $J_{i,j} = J_1 \delta_{|i - j|, 1}$, the ground state spontaneously breaks the $SU(2)$ symmetry of the model and develops magnetic LRO~\cite{Fradkin_2013}. The latter can be characterized by the spin structure factor,
\begin{equation}
S_L(\mathbf{q}) = \frac{1}{L^2} \sum_{i, j} \langle \hat{\mathbf{S}}_i \cdot \hat{\mathbf{S}}_j \rangle e^{i \mathbf{q} \cdot (\mathbf{r}_i - \mathbf{r}_j)},
\end{equation}
where $L\equiv L_x = L_y$ is the size of the lattice both in the $x$ and $y$ directions. In the zero magnetization sector, $M = \sum_i \langle \hat{S}^z_i \rangle = 0$, where $z$ has been chosen as the magnetization axis, the ground state corresponds to a N\'eel state. This is signaled by a non-zero value of the N\'eel order parameter $S^\text{N}_L \equiv S_L(\mathbf{q} = (\pi, \pi))$ in the thermodynamic limit, $S^\text{N}_\infty \equiv \lim_{L \to \infty} S^\text{N}_L \neq 0$. Note that, although the $SU(2)$ symmetry is broken, the ground state is still invariant under the subgroup $U(1)$, corresponding to rotations around the magnetization axis. 

\begin{figure}[t]
  \centering
  \includegraphics[width=1.0\linewidth]{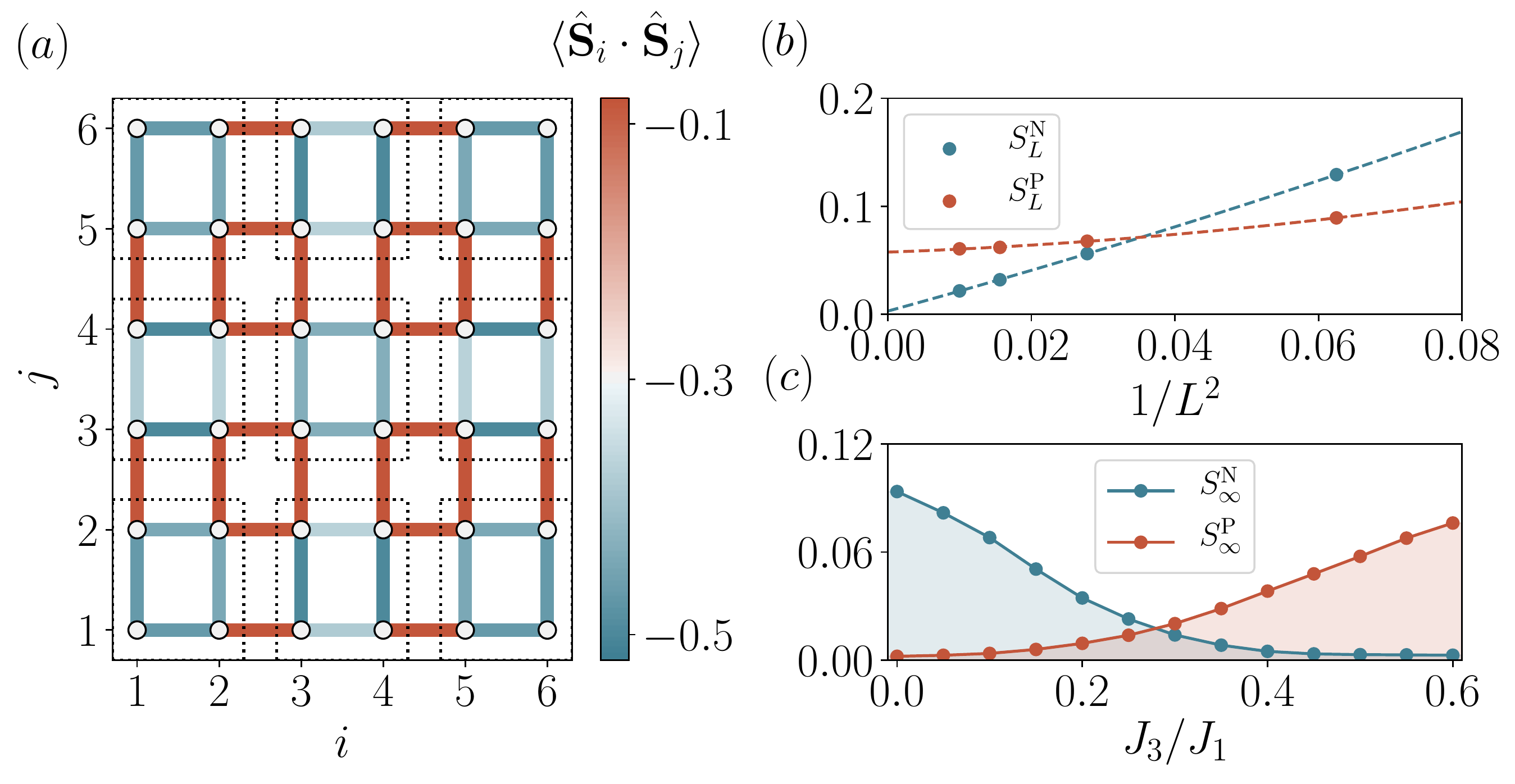}
\caption{\label{fig:PVBS} \textbf{Plaquette valence bond solid:}  {\bf (a)} Real-space bond pattern $\langle \vec{S}_i \cdot \vec{S}_j  \rangle$ for a lattice with $L^2 = 36$. For $J_3 / J_1 = 0.6$, the ground state belongs to a PBVS phase, characterized by a dimmerized bond pattern both in the $x$ and $y$ directions. The state breaks translational invariance, and a new four-site unit cell develops (marked by dotted squares in the figure). {\bf (b)} Finite-size scaling for the Neel ($S^\text{N}_L$) and PBVS ($S^\text{N}_L$) order parameters for $J_3 / J_1 = 0.5$. While the former vanishes in the thermodynamic limit, indicating a QP, the latter attains a finite value, associated with plaquette LRO. {\bf (c)} Thermodynamic limit extrapolation $S^\text{N/P}_\infty$ as a function of $J_3 / J_1$, indicating two distinct ordered phases.}
\end{figure}

Longer-range interactions beyond NN promote other types of magnetic LRO. Here we consider the $J_1-J_2-J_3$ Heisenberg model with interactions up to third-nearest-neighbors, this is, $J_{i,j} = J_1 \delta_{|i - j|, 1} + J_2 \delta_{|i - j|, \sqrt{2}} + J_3 \delta_{|i - j|, 2}$. For $J_3 = 0$, the ground state of the system is a collinear state at $\mathbf{q} = (0, \pi)$ or $\mathbf{q} = (\pi, 0)$ for $J_2 / J_1 \gtrsim 0.6$~\cite{Gong_2014}. At intermediate values around the classical transition point, $J_2 / J_1 = 0.5$, quantum fluctuations melt the magnetic order in a finite region of $J_2/J_1$, giving rise to a QP, with $\lim_{L \to \infty} S_L(\mathbf{q})=0$ $\forall \mathbf{q}$. Although the nature of the resulting phase is still disputed, density matrix renormalization group (DMRG) results point towards a PVBS phase~\cite{Gong_2014, Wang_2016}. This phase breaks translational invariance in the $x$ and $y$ directions, giving rise to a plaquette structure in the NN two-point correlators $\langle \hat{\mathbf{S}}_i \cdot \hat{\mathbf{S}}_j \rangle$, as shown in Fig.~\ref{fig:PVBS}{\bf (a)}. The plaquette LRO can be detected using the structure factor
\begin{equation}
S^\text{P}_L = \frac{1}{L^2}\sum_{\langle k,l \rangle} \epsilon(k,l)\left[\langle \hat{\mathbf{S}}_i \cdot \hat{\mathbf{S}}_j \, \hat{\mathbf{S}}_k \cdot \hat{\mathbf{S}}_l \rangle - \langle \hat{\mathbf{S}}_i \cdot \hat{\mathbf{S}}_j\rangle \langle \hat{\mathbf{S}}_k \cdot \hat{\mathbf{S}}_l \rangle\right],
\end{equation}
where the sum runs over NN pairs $\langle k, l \rangle$ and $\langle i, j \rangle$ is a fixed NN pair in the middle of the lattice. Finally, the form factor $\epsilon(k,l)$ is defined as in Ref.~\cite{Mambrini_2006}. The PVBS phase is thus characterized by $S^\text{P}_\infty \equiv \lim_{L\to\infty} S^\text{P}_L \neq 0$.

The PVBS can be further stabilized for $J_3\neq 0$. Using DMRG, it has been shown how this phase appears around the frustration line $(J_2 + J_3) / J_1 = 0.5$~\cite{Mambrini_2006}, where the classical energies of the magnetically-ordered states become equal. In the following, we focus on the case $J_2 = 0$, where the value of $S^\text{P}_\infty$ is larger~\cite{Mambrini_2006} and finite-size effects are expected to be less important. We use an MPS-based DMRG algorithm~\cite{Hauschild_2018} with bond dimension $D=1000$ to calculate the ground state of the frustrated Heisenberg model for various lattice sizes. In Fig.~\ref{fig:PVBS}{\bf (b)}, we show how the finite-size scaling of $S^\text{P}_L$ and $S^\text{N}_L$ for $J_3/J_1 = 0.5$ is consistent with a PBVS phase. In Fig.~\ref{fig:PVBS}{\bf (c)}, we show the thermodynamic limit extrapolation of the two order parameters, $S^\text{N}_\infty$ and $S^\text{P}_\infty$, as a function of $J_3 / J_1$. Although higher values of $L^2$ would be required to obtain the precise transition point, our results, consistent with previous works~\cite{Mambrini_2006, Murg_2009, Reuther_2011}, indicate the presence of two distinct quantum phases, a N\'eel and a plaquette-ordered phase.

\vspace{0.5ex}
\paragraph*{Higher-order topological quantum paramagnets.--} We now demonstrate how the PVBS phase described above presents, apart from LRO, non-trivial topological properties. In the last decades, a considerable effort has been devoted to the study of the frustrated Heisenberg model in a square lattice, in particular to uncover the nature of the QP phase. Here we claim that the PVBS is a HOSPT phase protected by a $U(1)\times C_4$ symmetry, this is, a combination of the conservation of the total magnetization $M$ and rotational invariance. To see this, we first notice how the bond pattern spontaneously generated in the PVBS state resembles the pattern required to create one of the earliest known examples of a HOSPT phase~\cite{Dubinkin_2019}. The latter corresponds to the ground state of a Heisenberg model with only NN interactions, where $J_1$ varies in space according to a plaquette-like structure, similarly to the one shown in Fig.~\ref{fig:PVBS}{\bf (a)}. In our case, however, such pattern arises in the bond correlators due to magnetic frustration via a SSB process starting from a translational invariant model.

\begin{figure}[t]
  \centering
  \includegraphics[width=1.0\linewidth]{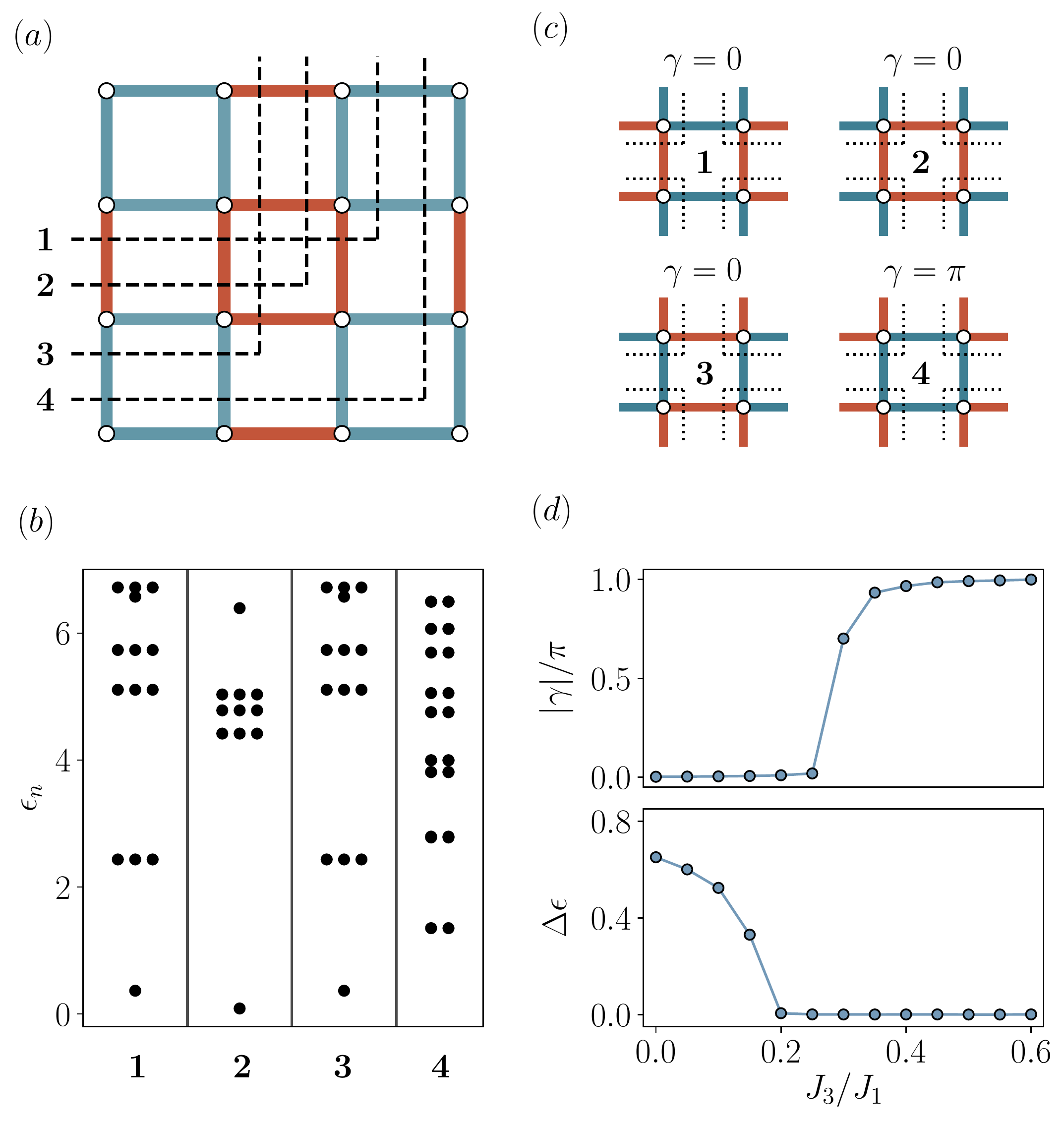}
\caption{\label{fig:HOSPT} \textbf{Higher-order symmetry-protected topology:}  {\bf (a)} In the figure, the dashed lines represent the four different types of bipartitions used to study the entanglement structure of the PVBS state. {\bf (b)} Lower part of the ES obtained from these bipartitions, for  $L^2 = 16$ and $J_3 / J_1 = 0.6$. The spectrum is doubly degenerate in one case, signaling its non-trivial higher-order topological nature. {\bf (c)} Local Berry phase $\gamma$ calculated at four different plaquettes in the middle of a square lattice with $L^2 = 100$ and for $J_3 / J_1 = 0.6$. A non-trivial phase of $\pi$ is obtained at the same plaquette where the ES shows degeneracy when cut through. {\bf (c)} Local Berry phase and ES degeneracy $\Delta \epsilon$ as a function of $J_3/J_1$ for a  $L^2 = 100$ lattice. The former is calculated at the central plaquette, while the latter is obtained by making a bipartition between one corner and the rest of the system.}
\end{figure}

To analyze the topological properties of the PVBS ground state, we first focus on the structure of its entanglement spectrum (ES)~\cite{Li_2008}. We define a bipartition of the system and write the ground state as $\ket{\psi_{\rm gs}} = \sum_n e^{-\epsilon_n / 2} \ket{\psi_n}_\text{A} \otimes \ket{\psi_n}_\text{B}$, where $\text{A}$ and $\text{B}$ are two subsystems, and $\{\epsilon_n\}$ is the ES, this is, the eigenenergies of the entanglement Hamiltonian $\hat{H}_\text{E}$, defined through the reduced density matrix $\rho_\text{A} = \text{Tr}_\text{B}\left(\ket{\Psi_{\rm gs}}\bra{\Psi_{\rm gs}}\right) = e^{-\hat{H}_\text{E}}$. The ES is degenerate for 1D topological phases~\cite{Pollmann_2010}, where cutting the chain into two halves creates a virtual edge, and this degeneracy is connected to the presence of localized states in the real boundaries. For 2D HOSPT phases, the same occurs when cutting the system such that a virtual corner is created, as we show in Fig.~\ref{fig:HOSPT}{\bf (a)}, which is associated in this case to the presence of corner states~\cite{Fukui_2018, You_2020, Dubinkin_2020}.

In Fig.~\ref{fig:HOSPT}{\bf (b)}, we show the lower part of the ES for four different bipartitions of a system with $L^2 = 16$ in the PVBS phase. The differences in the spectra stem from the inter or intra unit cell nature of the cut. At this point, it is important to note that the PBVS is four-fold degenerate as a consequence of the SSB. In a periodic or infinite system, the ground states are completely degenerate and they are only distinguished by the definition of the unit cell. The four different bipartitions on a finite system thus provide us with the same information as if we were cutting between unit cells for each of the four symmetry-broken ground states. Our results in Fig.~\ref{fig:HOSPT}{\bf (b)} prove that one of them ($\mathbf{4}$ in the figure) is topologically non trivial, and it is therefore expected to support localized corner states~\cite{Benalcazar_2017, Benalcazar_2019}. Due to finite-size effects, the lowest-energy state obtained using DMRG corresponds to configuration $\mathbf{2}$ in Fig.~\ref{fig:HOSPT} (see also Fig.~\ref{fig:PVBS}{\bf (a)}), and thus lacks corner states~\footnote{The ground state configuration that support corner states can be obtained numerically by adding a small perturbation that breaks the degeneracy and promotes it energetically. This, however, requires larger systems than those that could be accessed here with enough accuracy, where the energy difference between these states is smaller and so would be the required perturbation.}. However, we stress that the HOSPT nature of the phase is encoded in each state, which, as we will see, will lead to non-trivial topological phenomena.

To support our claim, we compute a local Berry phase $\gamma$ at different plaquettes for a system with $L^2 = 100$. The latter is a many-body topological invariant that is quantized in the presence of the $U(1)\times C_4$ symmetry to values $0$ and $\pi$ for trivial and HOSPT states, respectively~\cite{Araki_2020}. The results shown in Fig.~\ref{fig:HOSPT}{\bf (c)} for the different values of $\gamma$ coincide with those obtained for the ES, where we get $\gamma = \pi$ for the same plaquette that leads to degeneracy in the ES, and $\gamma = 0$ for the rest. Finally, we show both $\gamma$ and the ES degeneracy $\Delta \epsilon = \sum_n (-1)^n \epsilon_n$ for different values of $J_3 / J_1$. The latter is zero when the spectrum is degenerate and, although the transition points do not exactly match due to finite-size effects, the behavior of both quantities is consistent with a topological phase transition between a trivial magnetically-ordered state and a topological PVBS state. Finally, we note that this HOTQP phase can be regarded as a higher-order version of the Majumdar-Ghosh phase~\cite{Majumdar_1969a, Majumdar_1969b} found in the $J_1-J_2$ Heseinberg chain, which is a 1D VBS with a doubly-degenerate ground state, one of them being topologically non-trivial~\cite{Birnkammer_2020}.

\begin{figure}[t]
  \centering
  \includegraphics[width=1.0\linewidth]{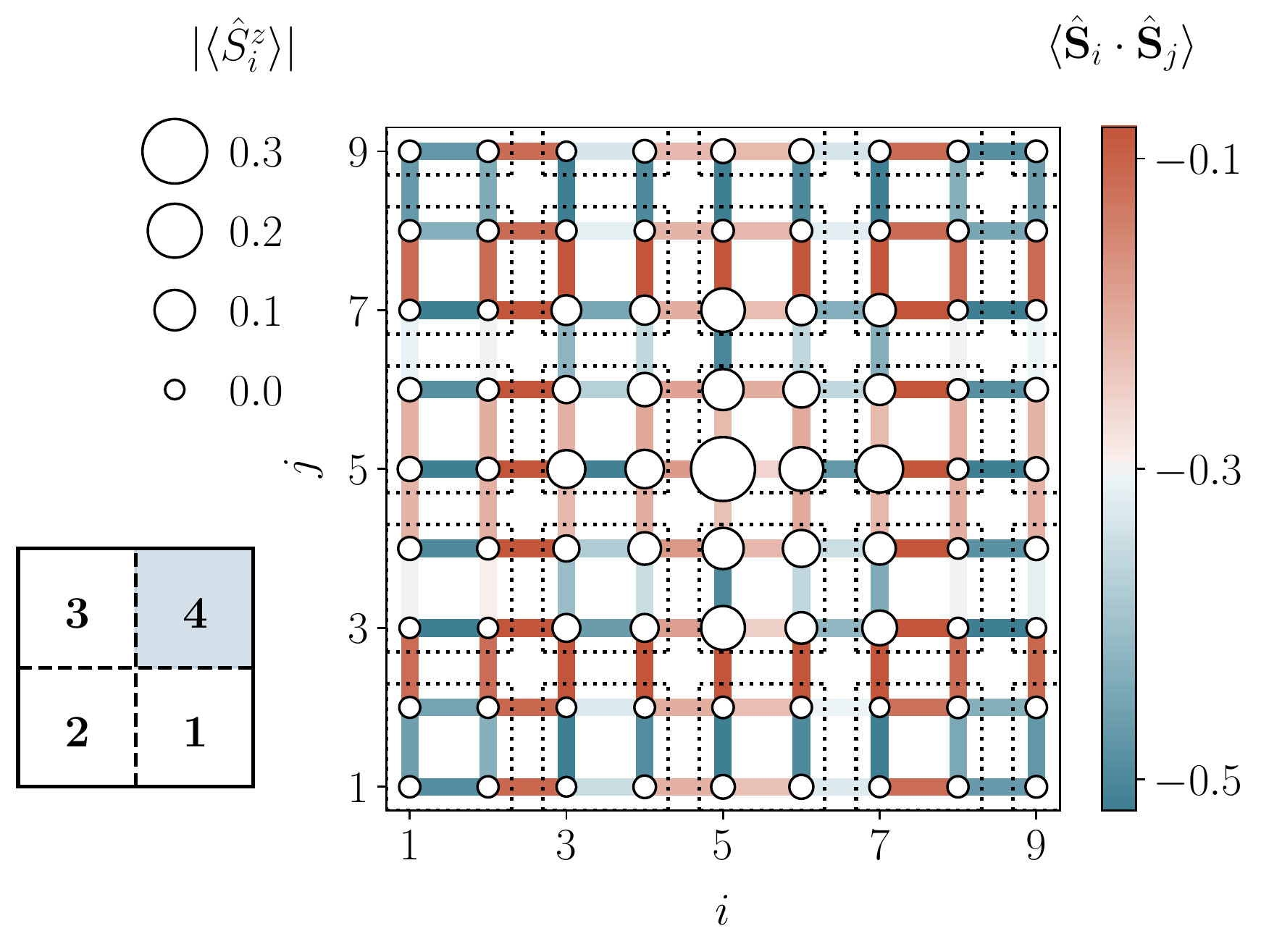}
\caption{\label{fig:defects} \textbf{Topological defects and corner-like states:}  Real-space bond pattern $\langle \hat{\mathbf{S}}_i \cdot \hat{\mathbf{S}}_j \rangle$ for a lattice with $L^2 = 81$ and $M = 1/2$. Two one-dimensional solitons are formed in the plaquette-ordered pattern, one vertical and one horizontal, crossing at the center of the lattice. The two defects divide the system into four parts that correspond to the four degenerate PVBS configurations of the ground state for $M = 0$, as indicated by the change in the unit cells (dotted squares) across the defects. One of them is topologically non-trivial, and the defects act as a corner in the bulk, separating it from topologically trivial regions. We present the net magnetization $|\langle S^z_i \rangle|$, which reveals a localized spin at this bulk corner.}
\end{figure} 

\vspace{0.5ex}
\paragraph*{Topological defects and corner-like states.--} The combination of LRO and higher-order topology in the HOTQP gives rise to novel strongly-correlated topological properties. To see this, we study the ground state of the frustrated Heisenberg model in a sector with non-zero total magnetization. In particular, we consider a lattice with $L^2 = 81$ and $M = 1/2$. Fig.~\ref{fig:defects} shows the real-space bond pattern for the ground state of the system in the HOTQP phase, where two topological defects are created in the plaquette-ordered structure, as compared to the homogeneous $M = 0$ case (Fig.~\ref{fig:PVBS}{\bf (a)}). The defects correspond to two 1D solitons that divide the system into four quadrants. Once the unit cell is defined, it becomes clear how each of the quadrants corresponds to one of the four degenerate ground-state configurations for $M = 0$ (Fig.~\ref{fig:HOSPT}{\bf (c)}). Since only one of them is topologically non-trivial, the point where the two defects cross corresponds to a corner between the latter and a trivial bulk. A localized spin is expected precisely at this point and this is confirmed by our results, shown in Fig.~\ref{fig:defects}, where we represent the local magnetization $\langle |\hat{S}^z_i| \rangle$. We note that, although topological defects have been recently considered for HOTIs, in particular disclinations~\cite{Benalcazar_2019, Liu_2019, Li_2020,Liu_2021, Peterson_2021} and dislocations~\cite{Queiroz_2019, Roy_2020}, they correspond to static defects imposed externally. Our results show for the first time how localized corner-like states can emerge bound to dynamical topological defects in a translational invariant system. The latter can be thought as the higher-order 2D version of the localized states induced by solitons in the Su-Schrieffer-Heeger model~\cite{Su_1979}, recently generalized to the bosonic case~\cite{Gonzalez-Cuadra_2019a, Gonzalez-Cuadra_2019b, Gonzalez-Cuadra_2020a, Gonzalez-Cuadra_2020b}, and the Majumdar-Ghosh phase~\cite{Birnkammer_2020}.

\vspace{0.5ex}
\paragraph*{Quantum simulation with ultracold molecules.--} Although the frustrated Heisenberg model describes a variety of materials, observing a HOTQP can be challenging in solid-state devices. We now discuss the possibility of preparing this phase using quantum simulators based on ultracold molecules. Consider in particular a 2D array of polar molecules in an optical lattice, whose motional degree of freedom is frozen. In this situation, the molecules can be regarded as quantum rotors interacting via dipolar interactions~\cite{Micheli_2006, Barnett_2006, Gorshkov_2011a, Gorshkov_2011b}. One can then choose two rotational states $\ket{J,M}$ to build spin operators, where $J$ and $M$ are the total and third component of the angular momentum operator $\hat{\mathbf{J}}$. For a strong enough external electric field aligned perpendicular to the lattice plane, and by taking $\ket{\uparrow} = \ket{1, 0}$ and $\ket{\downarrow} = \ket{0, 0}$, the system is described by the effective Hamiltonian~\cite{Yao_2018}
\begin{equation}
\hat{H}_\text{eff} = \sum_{i,j} \frac{g}{|i - j|^3} \left[2d^2_{00}\left(\hat{S}^x_i \hat{S}^x_j + \hat{S}^y_i \hat{S}^y_j\right) + (\mu_0 - d_0)^2 \hat{S}^z_i \hat{S}^z_j\right],
\end{equation}
with $g=1/(4\pi\epsilon_0)$, $d_{00} =\langle \uparrow | \hat{d}_z | \downarrow \rangle$, $d_0 =\langle \downarrow | \hat{d}_z | \downarrow \rangle$ and $\mu_0 =\langle \uparrow | \hat{d}_z | \uparrow \rangle$, where $\hat{d}_z$ is the third component of the dipole operator $\hat{\mathbf{d}}$. By tunning the electric field, the dipolar Heisenberg point can be reached at $2d^2_{00} = (\mu_0 - d_0)^2$~\cite{Yao_2018}, corresponding to Hamiltonian~\eqref{eq:model} with $J_{i,j} = 2g\,d^2_{00} / |i - j|^3$. In this case, the first three largest interaction terms are related by $(J_2 + J_3) / J_1 = 1 / 2^{3/2} + 1 / 2^3 \approx 0.479$, a value that is close to the frustration point $0.5$. It is therefore natural to ask whether the ground state of the dipolar Heisenberg model in a square lattice is a HOTQP.

\begin{figure}[t]
  \centering
  \includegraphics[width=1.0\linewidth]{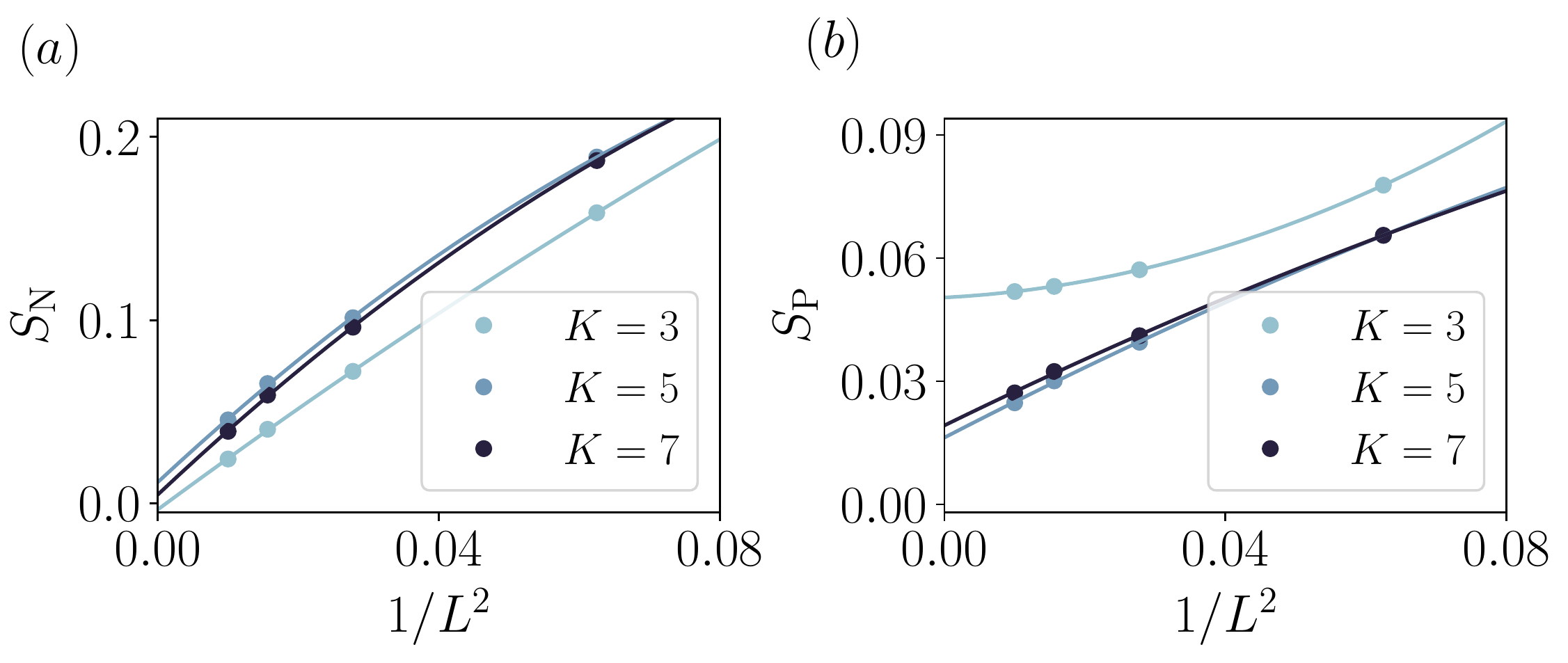}
\caption{\label{fig:dipolar} \textbf{Dipolar Heisenberg model:}  Finite-size scaling of the order parameters $S^\text{N}_L$ and $S^\text{P}_L$ in the ground state of the dipolar Heisenberg model for different truncations $K$ of the dipolar interactions (see main text). The results are consistent with a PVBS.}
\end{figure} 

Previous works have studied the model using various approximations in the presence of anisotropy, but have found no sign of a QP in the zero-anisotropy limit~\cite{Zou_2017, Keles_2018a}. Our DMRG results, however, show that the N\'eel order parameter vanishes in the ground state of the model. In Fig.~\ref{fig:dipolar}{\bf (a)}, we show the finite-size scaling of $S^{\text N}_L$ for different truncations of the dipolar interactions, this is, by keeping only the $K$ largest terms. In every case the results are consistent with a QP and for $K>5$ the values do not change noticeably. Moreover, the results for the finite-size scaling of $S^{\text P}_L$ are consistent with a PVBS (Fig.~\ref{fig:dipolar}{\bf (b)}) for every studied value of $K$. Even if the value of  $S^{\text P}_\infty$ is reduced from $K=3$ to $K=5$, it remains non-zero and it does not change noticeable for $K>5$. Finally, we have calculated the local Berry phase in the ground state of the dipolar Heisenberg model, obtaining a value of $\gamma = 0.97 \pi$ for the same plaquette as in Fig.~\ref{fig:HOSPT}{\bf (c)}, where a perfect quantization is only lacking due to finite-size effects.

Our results therefore indicate that a HOTQP could be prepared using ultracold molecules in an optical lattice. The real-space plaquette order structure, as well as the presence of topological defects and corner-like states, could be revealed using a quantum gas microscope with single-site resolution~\cite{Bakr_2009, Sherson_2010, Covey_2018}. The non-trivial topological properties could be accessed by measuring the entanglement spectrum, which can be achieved in an efficient manner using near-term quantum simulators~\cite{Kokail_2021a, Kokail_2021b}. Although here we focus on ultracold molecules, we notice that similar physics could be also simulated using magnetic atoms~\cite{Lahaye_2009}, where spin models with dipolar interactions have already been realized experimentally~\cite{dePaz_2013, dePaz_2016}, or trapped ions~\cite{Porras_2011, Grass_2014, Bermudez_2017, Monroe_2021}.

\vspace{0.5ex}
\paragraph*{Conclusions and outlook.--} In this work, we investigated how magnetic frustration can induce non-trivial higher-order topological properties. In particular, we studied the well-known PVBS in the frustrated Heisenberg model and we showed how it belongs to a HOTQP phase. Although we focus on the $J_1-J_3$ model, our results could be generalized to any PVBS, in particular to the ground state of the $J_1-J_2$ model, since they are adiabatically connected. Moreover, we revealed how the interplay between LRO and higher-order topology can give rise to new topological excitations, such as corner-like states in the bulk bound to dynamical topological defects.

Our results show that spin liquids are not the only QPs with interesting topological properties, and that a HOSPT phase was hiding in plain sight in paradigmatic spin models. Moreover, we show how this phase is stabilized by dipolar interactions, indicating how the physics of HOTQP states could be further explored in atomic quantum simulators, which offer much higher control compared to natural materials. In particular, experiments with ultracold molecules, magnetic atoms or trapped ions could achieve larger system sizes than those accessible using classical simulations, which could facilitate for instance the study of the topological phase transition that gives rise to the HOTQP. Finally, the role of vortex defects in the N\'eel to PVBS phase transition was recently studied in the $J_1-J_2$ model~\cite{You_2020b}. The former are examples of fracton excitations~\cite{Nandkishore_2019}, and their condensation could explain the presence of deconfined critical points~\cite{Senthil_2004,Wang_2016}. Therefore, it would be interesting to analyze the role of the solitons considered here to further elucidate the nature of the corresponding topological critical point, as well as to find other mechanisms that can give rise to HOSPT phases induced by SSB.

\vspace{0.5ex}
\paragraph*{Acknowledgements.--} We thank L. Barbiero, A. Bermudez, A. Dauphin, J. Fraxanet, S. Juli\`a and M. Lewenstein for useful discussions. We acknowledge support from ERC AdG NOQIA, Agencia Estatal de Investigaci\'on (``Severo Ochoa'' Center of Excellence CEX2019-000910-S, Plan National FIDEUA PID2019-106901GB-I00/10.13039/501100011033, FPI), Fundaci\'o Privada Cellex, Fundaci\'o Mir-Puig, and from Generalitat de Catalunya (AGAUR Grant No. 2017 SGR 1341, CERCA program, QuantumCAT \_U16-011424, co-funded by ERDF Operational Program of Catalonia 2014-2020), MINECO-EU QUANTERA MAQS (funded by State Research Agency (AEI) PCI2019-111828-2 / 10.13039/501100011033), EU Horizon 2020 FET-OPEN OPTOLogic (Grant No 899794), and the National Science Centre, Poland-Symfonia Grant No. 2016/20/W/ST4/00314.

\bibliography{bibliography}

\end{document}